\documentclass[aps,prl,preprint,tightenlines,superscriptaddress,showpacs,byrevtex]{revtex4}

\usepackage{graphicx} 
\usepackage{dcolumn}  

\graphicspath{{ps}}

\newcommand{\ycp}{$y_{CP}$}
\newcommand{\zdmix}{$D^0$-$\overline{D}^0$}
\newcommand{\zdecay}{$D^0\to K_S^0K^+K^-$}
\newcommand{\zdtau}{$\Delta_{\tau}$}

\begin{document}

\preprint{\vbox{ \hbox{   }
                 \hbox{BELLE-CONF-0846}
}}

\title{ \quad\\[0.5cm]  Measurement of \ycp\ in $D$ meson decays to $CP$ eigenstates}

\affiliation{Budker Institute of Nuclear Physics, Novosibirsk}
\affiliation{Chiba University, Chiba}
\affiliation{University of Cincinnati, Cincinnati, Ohio 45221}
\affiliation{Department of Physics, Fu Jen Catholic University, Taipei}
\affiliation{Justus-Liebig-Universit\"at Gie\ss{}en, Gie\ss{}en}
\affiliation{The Graduate University for Advanced Studies, Hayama}
\affiliation{Gyeongsang National University, Chinju}
\affiliation{Hanyang University, Seoul}
\affiliation{University of Hawaii, Honolulu, Hawaii 96822}
\affiliation{High Energy Accelerator Research Organization (KEK), Tsukuba}
\affiliation{Hiroshima Institute of Technology, Hiroshima}
\affiliation{University of Illinois at Urbana-Champaign, Urbana, Illinois 61801}
\affiliation{Institute of High Energy Physics, Chinese Academy of Sciences, Beijing}
\affiliation{Institute of High Energy Physics, Vienna}
\affiliation{Institute of High Energy Physics, Protvino}
\affiliation{Institute for Theoretical and Experimental Physics, Moscow}
\affiliation{J. Stefan Institute, Ljubljana}
\affiliation{Kanagawa University, Yokohama}
\affiliation{Korea University, Seoul}
\affiliation{Kyoto University, Kyoto}
\affiliation{Kyungpook National University, Taegu}
\affiliation{\'Ecole Polytechnique F\'ed\'erale de Lausanne (EPFL), Lausanne}
\affiliation{Faculty of Mathematics and Physics, University of Ljubljana, Ljubljana}
\affiliation{University of Maribor, Maribor}
\affiliation{University of Melbourne, School of Physics, Victoria 3010}
\affiliation{Nagoya University, Nagoya}
\affiliation{Nara Women's University, Nara}
\affiliation{National Central University, Chung-li}
\affiliation{National United University, Miao Li}
\affiliation{Department of Physics, National Taiwan University, Taipei}
\affiliation{H. Niewodniczanski Institute of Nuclear Physics, Krakow}
\affiliation{Nippon Dental University, Niigata}
\affiliation{Niigata University, Niigata}
\affiliation{University of Nova Gorica, Nova Gorica}
\affiliation{Osaka City University, Osaka}
\affiliation{Osaka University, Osaka}
\affiliation{Panjab University, Chandigarh}
\affiliation{Peking University, Beijing}
\affiliation{Princeton University, Princeton, New Jersey 08544}
\affiliation{RIKEN BNL Research Center, Upton, New York 11973}
\affiliation{Saga University, Saga}
\affiliation{University of Science and Technology of China, Hefei}
\affiliation{Seoul National University, Seoul}
\affiliation{Shinshu University, Nagano}
\affiliation{Sungkyunkwan University, Suwon}
\affiliation{University of Sydney, Sydney, New South Wales}
\affiliation{Tata Institute of Fundamental Research, Mumbai}
\affiliation{Toho University, Funabashi}
\affiliation{Tohoku Gakuin University, Tagajo}
\affiliation{Tohoku University, Sendai}
\affiliation{Department of Physics, University of Tokyo, Tokyo}
\affiliation{Tokyo Institute of Technology, Tokyo}
\affiliation{Tokyo Metropolitan University, Tokyo}
\affiliation{Tokyo University of Agriculture and Technology, Tokyo}
\affiliation{Toyama National College of Maritime Technology, Toyama}
\affiliation{Virginia Polytechnic Institute and State University, Blacksburg, Virginia 24061}
\affiliation{Yonsei University, Seoul}
  \author{I.~Adachi}\affiliation{High Energy Accelerator Research Organization (KEK), Tsukuba} 
  \author{H.~Aihara}\affiliation{Department of Physics, University of Tokyo, Tokyo} 
  \author{D.~Anipko}\affiliation{Budker Institute of Nuclear Physics, Novosibirsk} 
  \author{K.~Arinstein}\affiliation{Budker Institute of Nuclear Physics, Novosibirsk} 
  \author{T.~Aso}\affiliation{Toyama National College of Maritime Technology, Toyama} 
  \author{V.~Aulchenko}\affiliation{Budker Institute of Nuclear Physics, Novosibirsk} 
  \author{T.~Aushev}\affiliation{\'Ecole Polytechnique F\'ed\'erale de Lausanne (EPFL), Lausanne}\affiliation{Institute for Theoretical and Experimental Physics, Moscow} 
  \author{T.~Aziz}\affiliation{Tata Institute of Fundamental Research, Mumbai} 
  \author{S.~Bahinipati}\affiliation{University of Cincinnati, Cincinnati, Ohio 45221} 
  \author{A.~M.~Bakich}\affiliation{University of Sydney, Sydney, New South Wales} 
% ...
\collaboration{The Belle Collaboration}

\noaffiliation

\begin{abstract}
We present a measurement of the \zdmix\ mixing parameter \ycp\ using a flavor-untagged sample of \zdecay\ decays. The measurement is based on a 673 fb$^{-1}$ data sample recorded by the Belle detector at the KEKB asymmetric-energy $e^+ e^-$ collider. We find $y_{CP} = (0.21\pm 0.63 ({\rm stat.})\pm 0.78 (\rm syst.) \pm 0.01(\rm model))\%$.
\end{abstract}


\maketitle

\tighten

{\renewcommand{\thefootnote}{\fnsymbol{footnote}}}
\setcounter{footnote}{0}

Particle-antiparticle mixing has been observed in several systems of neutral mesons: neutral kaons, $B_d$ and $B_s$ mesons.
As in the kaon and $B$-meson systems, the $D^0$ - $\overline{D}^0$ are produced in flavor eigenstates. The mixing occurs through weak interactions between the quarks and gives rise to two different mass eigenstates $|D_{1,2}\!>=p|D^0\!\!>\pm~ q|\bar{D}^0\!\!>$, where $p$ and $q$ are complex coefficients satisfying $|p|^2+|q|^2=1$. The time evolution of flavor eigenstates, $D^0$ and $\overline{D}^0$, is governed by the mixing parameters
$x = (m_1-m_2)/\Gamma$ and $y = (\Gamma_1-\Gamma_2)/2\Gamma$, where $m_{1,2}$ and $\Gamma_{1,2}$ are the masses and widths of the two mass eigenstates $D_{1,2}$, and
$\Gamma = (\Gamma_1+\Gamma_2)/2$.
In the Standard Model (SM), $D^0$ - $\overline{D}^0$ mixing is strongly GIM suppressed for $d$ and $s$ quarks and CKM suppressed for $b$ quark box diagrams, 
and is dominated by long distance effects \cite{mix_th}. 
As the mixing rate is expected to be small within the SM, it is sensitive to the contribution of new, as yet unobserved processes and particles. 
The largest SM predictions for the parameters $x$ and $y$, which include the impact of long distance dynamics, are of order $1$\% \cite{mix_th}.
Various $D^0$ decay modes have been used to measure or constrain $x$ and $y$ \cite{PDGAsner}. Evidence for \zdmix\ has been found in $D^0\to K^+K^-/\pi^+\pi^-$ \cite{Staric,AubertMix}, $D^0\to K^+\pi^-$ \cite{Aubert1,CDF} and $D^0\to K^+\pi^-\pi^0$ \cite{kpipi0} decays. The world average \cite{hfag,Schwartz} of $D^0$ mixing parameter \ycp\ measured in $D^0\to K^+K^-/\pi^+\pi^-$ decays is $y_{CP} = (1.132 \pm 0.266 )\%$, where $y_{CP}=y$ if $CP$ is conserved. Here we study the self-conjugate decay \zdecay\ \cite{CC}.

\begin{figure}[t]
 \includegraphics[height=4.5cm]{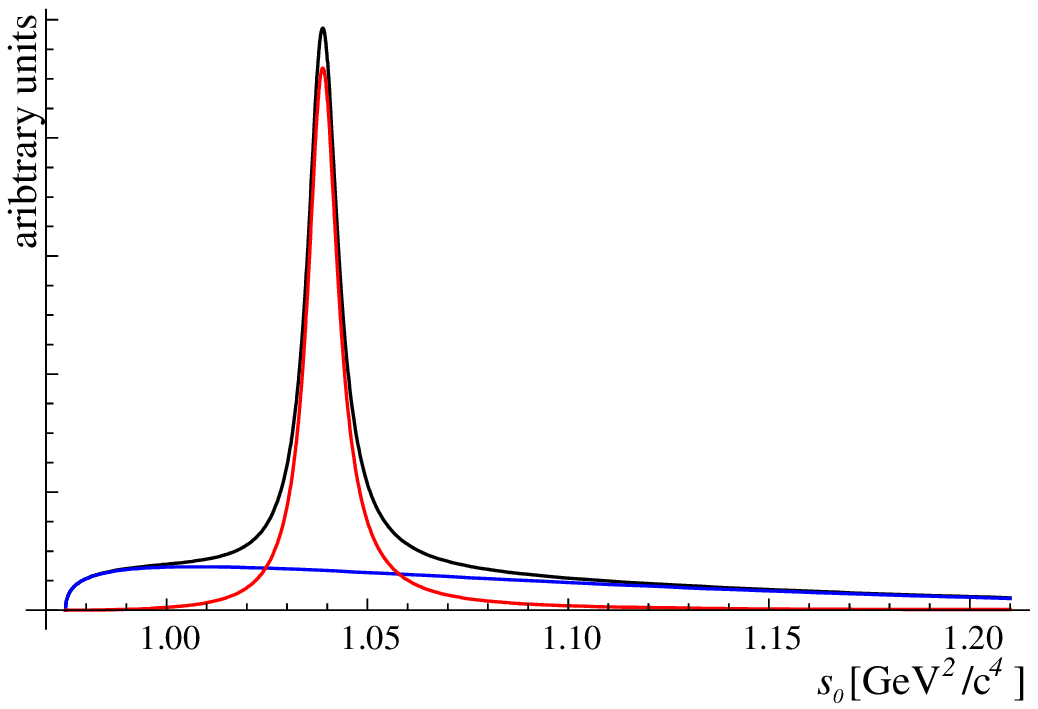}
 \includegraphics[height=4.5cm]{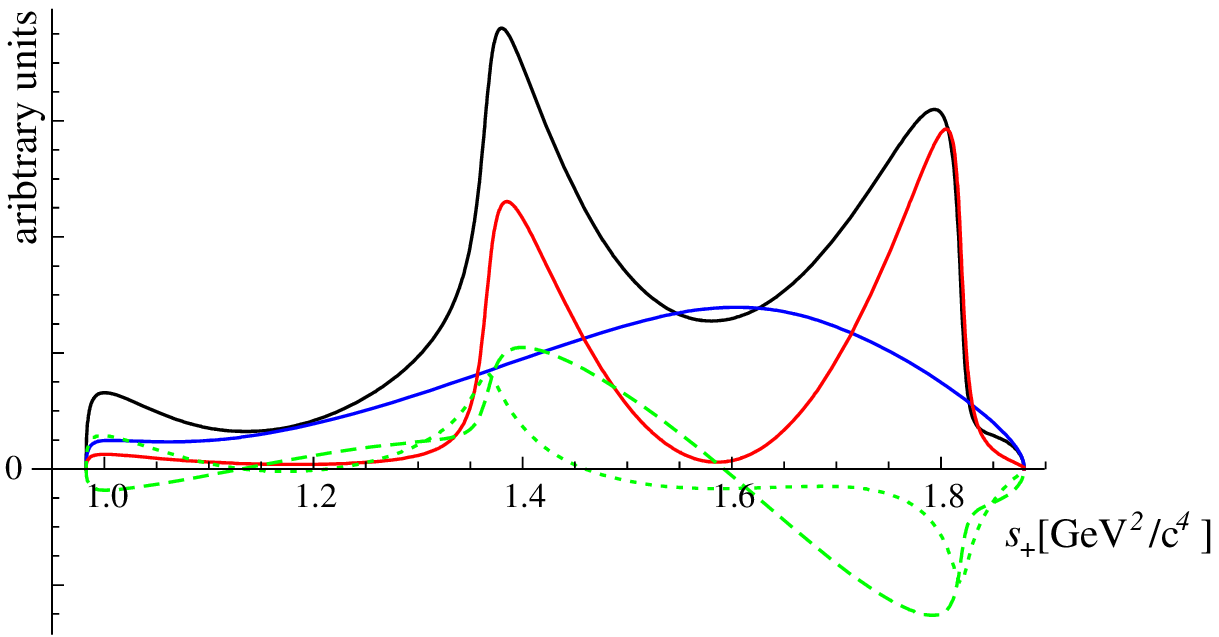}
 \caption{$s_0$ (left) and $s_+$ (right) Dalitz plot projections of $|{\cal M}(s_0,s_+)|^2$ (black line), $|{\cal A}_1(s_0,s_+)|^2$ (blue line ), $|{\cal A}_2(s_0,s_+)|^2$ (red line), $2Re[{\cal A}_1(s_0,s_+){\cal A}_2^{\ast}(s_0,s_+)|]$ (green dotted line) and $2Im[{\cal A}_1(s_0,s_+){\cal A}_2^{\ast}(s_0,s_+)|]$ (green dashed line) for Dalitz model given in \cite{BaBar}.}	
 \label{fig_DalitzModel}
\end{figure}

The time dependent decay rate of an initially produced $D^0$ or $\overline{D}^0$ can be expressed as \cite{LiMing,AsnerCleo}
\begin{eqnarray}
 |{\cal M}(s_0,s_+,t)|^2 & = & |{\cal A}_1(s_0,s_+)|^2e^{-\frac{t(1+y)}{\tau}} + |{\cal A}_2(s_0,s_+)|^2e^{-\frac{t(1-y)}{\tau}} \nonumber\\
& & + 2Re[{\cal A}_1(s_0,s_+){\cal A}_2^{\ast}(s_0,s_+)]cos(\frac{xt}{\tau})e^{-\frac{t}{\tau}} \nonumber\\
& & + 2Im[{\cal A}_1(s_0,s_+){\cal A}_2^{\ast}(s_0,s_+)]sin(\frac{xt}{\tau})e^{-\frac{t}{\tau}} \label{eq_mD0}\\
|\overline{{\cal M}}(s_0,s_+,t)|^2 & = &  |\overline{{\cal A}_1}(s_0,s_+)|^2e^{-\frac{t(1+y)}{\tau}} + |\overline{{\cal A}_2}(s_0,s_+)|^2e^{-\frac{t(1-y)}{\tau}}\nonumber \\
 &   &  + 2Re[\overline{{\cal A}_1}(s_0,s_+)\overline{{\cal A}_2}^{\ast}(s_0,s_+)]cos(\frac{xt}{\tau})e^{-\frac{t}{\tau}} \nonumber\\
 &   &  + 2Im[\overline{{\cal A}_1}(s_0,s_+)\overline{{\cal A}_2}^{\ast}(s_0,s_+)]sin(\frac{xt}{\tau})e^{-\frac{t}{\tau}} \label{eq_mD0bar},
\end{eqnarray}
where $\tau=1/\Gamma$ is the $D^0$ lifetime, $s_0$ and $s_+$ are invariant masses squared of $K^+K^-$ and $K_SK^+$ pairs, respectively.
The decay amplitudes ${\cal A}_{1}$ and ${\cal A}_{2}$ can be expressed with $D^0$ and $\overline{D}^0$ decay amplitudes ${\cal A}$ and $\overline{\cal A}$ as
\begin{eqnarray}
 {\cal A}(s_0,s_+)            & = & \sum_r a_r e^{i\phi_r}{\cal A}_r(s_0,s_+)\\
 \overline{\cal A}(s_0,s_+)   & = & \sum_r \overline{a}_r e^{i\overline{\phi}_r}\overline{\cal A}_r(s_0,s_+)\\
 {\cal A}_1(s_0,s_+) & = & \frac{1}{2}\left ({\cal A}(s_0,s_+) + \overline{\cal A}(s_0,s_+) \right ) = \sum~\mbox{$CP=+1$ and flavor eigenstates}\\
 {\cal A}_2(s_0,s_+) & = & \frac{1}{2}\left ({\cal A}(s_0,s_+) - \overline{\cal A}(s_0,s_+) \right ) = \sum~\mbox{$CP=-1$ and flavor eigenstates},
\end{eqnarray}
where ${\cal A}$ and $\overline{\cal A}$ are summed over resonant contributions $r$ found in \zdecay\ decays. 
In the limit of $CP$ conservation $a_r=\overline{a_r}$, $\phi_r=\overline{\phi}_r$ and $\overline{\cal A}(s_0,s_+)={\cal A}(s_0,s_-)$.
The existing Dalitz plot analyses of \zdecay decays \cite{BaBarII,BaBar} observed contribution of $CP$ ($K^0_S a_0(980)^0$, $K^0_S \phi(1020)$, 
$K^0_S f_0(1370)$, $K^0_S f_2(1270)$, $K^0_S a_0(1450)^0$, $K^0_S f_0(980)$), Cabbibo-allowed ($K^- a_0(980)^+$, $K^- a_0(1450)^+$) and doubly Cabbibo-suppressed 
($K^+ a_0(980)^-$) flavor eigenstates. Figure \ref{fig_DalitzModel} shows time integrated $s_0$ and $s_+$ projections of  $|{\cal M}(s_0,s_+)|^2$, 
$|{\cal A}_1(s_0,s_+)|^2$, $|{\cal A}_2(s_0,s_+)|^2$, $2Re[{\cal A}_1(s_0,s_+){\cal A}_2^{\ast}(s_0,s_+)|]$ and $2Im[{\cal A}_1(s_0,s_+){\cal A}_2^{\ast}(s_0,s_+)|]$ obtained by Dalitz model given in Ref. \cite{BaBar}. The integral of $2Re[{\cal A}_1(s_0,s_+){\cal A}_2^{\ast}(s_0,s_+)|]$ and $2Im[{\cal A}_1(s_0,s_+){\cal A}_2^{\ast}(s_0,s_+)|]$ over $s_+$ yields 0.

The $|{\cal A}_1|^2$ and $|{\cal A}_2|^2$ parts of the decay rate have different time dependence (Eq. \ref{eq_mD0} and \ref{eq_mD0bar}) and also very different dependence in the $s_0$ (Fig. \ref{fig_DalitzModel} (left)). In any given $s_0$ region the lifetime of $D^0$ candidates is given by
\begin{equation}
 \tau'=f_1\frac{\tau}{1+y_{CP}}+(1-f_1)\frac{\tau}{1-y_{CP}},
\end{equation}
where $\tau$ is the mean $D^0$ lifetime $1/\Gamma$, $f_1 = \oint |{\cal A}_1|^2/\oint (|{\cal A}_1|^2+|{\cal A}_2|^2)$ and $CP$ conservation is assumed.
The lifetime difference of $D^0$ candidates in two different regions is then proportional to the mixing parameter $y_{CP}$
\begin{equation}
 \Delta_{\tau}= \frac{\tau'-\tau''}{\tau'+\tau''}=y_{CP}\frac{f''_1-f'_1}{1+y_{CP}(1-f''_1-f'_1)}\approx y_{CP}(f''_1-f'_1).
 \label{eq_deltatau}
\end{equation}
The best $m(K^+K^-)$ intervals from which $D^0$ lifetimes are measured and compared are those that minimize the statistical uncertainty on $y_{CP}$ and are found to be: region around $\phi(1020)$ peak $m(K^+K^-) \in [1.015, 1.025]$ GeV$/c^2$ (denoted as ON) and intervals $m(K^+K^-) \in [2m_{K^{\pm}},1.010]$ GeV$/c^2$ and 
$m(K^+K^-) \in [1.033,1.100]$ GeV$/c^2$ (the union of this two intervals is denoted as OFF), where $m_{K^{\pm}}$ is the nominal $K^{\pm}$ mass.

The data were recorded by the Belle detector at the KEKB asymmetric-energy collider \cite{KEKB}. 
The Belle detector is a large-solid-angle magnetic spectrometer that consists of a silicon vertex detector (SVD), a 50-layer central drift chamber (CDC), an array of
aerogel threshold Cherenkov counters (ACC), a barrel-like arrangement of time-of-flight scintillation counters (TOF), and an electromagnetic calorimeter
(ECL) comprised of CsI(Tl) crystals located inside a superconducting solenoid coil that provides a 1.5~T magnetic field.  An iron flux-return located outside of
the coil is instrumented to detect $K_L^0$ mesons and to identify muons (KLM). The detector is described in detail elsewhere~\cite{Belle}.
Two inner detector configurations were used. A 2.0 cm beampipe and a 3-layer silicon vertex detector was used for the first sample
of 156 fb$^{-1}$, while a 1.5 cm beampipe, a 4-layer silicon detector and a small-cell inner drift chamber were used to record  
the remaining 517 fb$^{-1}$ of data.

The $K_S^0$ candidates are reconstructed in the $\pi^+\pi^-$ final state; we require that the pion candidates form a common vertex at least 0.9~mm from the  $e^+e^-$ interaction point (IP) in plane perpendicular to the beam axis and have an invariant mass within $\pm 30$~MeV/$c^2$ of $K_S^0$ nominal mass. We reconstruct $D^0$ candidates by combining the $K_S^0$ candidate with two oppositely charged tracks assigned as kaons. These tracks are required to have at least one SVD hit in both $r-\phi$ and $z$ coordinates. A $D^0$ momentum greater than 2.55~GeV/$c$ in the $e^+e^-$ center-of-mass (CM) frame is required to reject $D$ mesons produced in $B$ mesons decays and to suppress combinatorial background.

The decay point of $D^0$ candidate is determined by refitting one of the charged kaons and $K^0_S$ candidate to a common vertex \cite{vertex}; confidence levels exceeding $10^{-3}$ are required for the both fits. Out of two possibilities the one with lowest $\chi^2$ value of the fit is used. In addition we require that $K_S^0K^+K^-$ and $K^+K^-$ combinations originate from the common vertex by rejecting candidates of this two fits  with confidence levels lower than $10^{-3}$. The $D^0$ production point is taken to be the intersection of the $D^0$ momentum vector with the IP. The proper decay time of the $D^0$ candidate is then calculated from the projection of the vector joining the production and decay points, $\vec{L}$, onto the $D^0$ momentum vector, $t=(m_{D^0}/p_D)\vec{L}\cdot(\vec{p}_D/p_D)$, where $m_{D^0}$ is the nominal $D^0$ mass. The decay time uncertainty $\sigma_t$ is evaluated event-by-event, and we require $\sigma_t<600$ fs (the maximum of $\sigma_t$ distribution is at $\sim230$~fs).

\begin{figure}[t!]
 \begin{center}
 \includegraphics[width=0.8\textwidth]{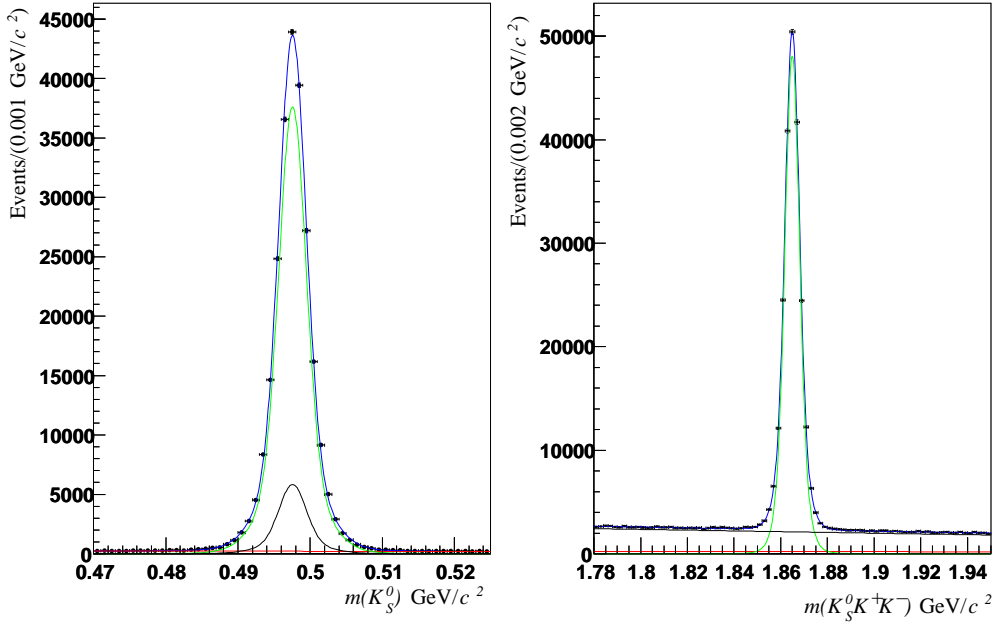}
 \end{center}
 \caption{The distribution of $m(K_S)$ with $m(K^0_SK^+K^-)\in [1.85,1.88]$~GeV/$c^2$ (left) and $m(K^0_SK^+K^-)$ with $m(K_S)\in [0.490,0.505]$~GeV/$c^2$ (right). Superimposed on the data (points with error bars) are projections of the $m(K_S)-m(K^0_SK^+K^-)$ fit (result from the fit (solid blue lines), signal contribution (solid green line), true $K^0_S$ (solid black line) and rest of the background (solid red line)).}
 \label{fig_md0mksfit_untagged_data}
\end{figure}
The signal and background yields are determined from a two-dimensional fit to the invariant masses of $K^0_S$ and $D^0$ candidates. According to Monte Carlo (MC) simulated distributions of $m(K^0_S)$ and $m(K^0_SK^+K^-)$, events can be divided into three categories: (1) signal \zdecay decays; (2) true $K^0_S$ candidates combined with random charged kaons (one or both);  and (3) rest of the background. We parametrize the signal shape by a sum of a three two-dimensional-Gaussian function and a product of two one-dimensional-Gaussian functions used to describe long tails in both variables (the contribution of the latter is small, $\sim0.1\%$). The second category is described by a sum of three Gaussians for $m(K^0_S)$  and a linear function for  $m(K^0_SK^+K^-)$. The third category is described by a product of the linear functions. The $m(K^0_S)$ and $m(K^0_SK^+K^-)$ distributions are shown in Fig. \ref{fig_md0mksfit_untagged_data} along with the projections of the fit result. 
The MC simulation shows that a small fraction ($\sim 0.1\%$) of events are decays of $D^0$ mesons to $K^+K^-\pi^+\pi^-$ final state (charged pions do not originate from $K^0_S$ decay). These events are peaking in $m(K^0_SK^+K^-)$, but not in $m(K^0_S)$. The projections of $m(K^0_SK^+K^-)$ for events in $m(K^0_S)$ sidebands are checked for possible contribution of $D^0\to K^+K^-\pi^+\pi^-$ decays. We find no contribution of $D^0\to K^+K^-\pi^+\pi^-$ decays.
The fit is performed to obtain scaling factors for the background fractions, and then tune them in the MC event-by-event in order to achieve better agreement in $m(K^0_S)$ and $m(K^0_SK^+K^-)$ distributions between MC and data events.

The sample of events for the lifetime measurement is selected using $|m'(K^0_S)|$ and $|m'(K^0_SK^+K^-)|$, where $m'(K^0_S)$ and $m'(K^0_SK^+K^-)$ are rotated $K^0_S$ and $D^0$ candidate masses according to
\begin{eqnarray}
 m'(K^0_S) & = & \frac{m(K^0_S)-m_{K^0_S}}{\sigma(K^0_S)}\\
 m'(K^0_SK^+K^-) & = & \rho\frac{m(K^0_S)-m_{K^0_S}}{\sigma(K^0_S)\sqrt{1-\rho^2}}-\frac{m(K^0_SK^+K^-)-m_{D^0}}{\sigma(K^0_SK^+K^-)\sqrt{1-\rho^2}},
\end{eqnarray}
where $m_{K^0_S}=497.57\pm 0.01$~MeV$/c^2$ and $m_{D^0}=1864.96\pm 0.01$~MeV$/c^2$ are fitted $K^0_S$ and $D^0$ masses, $\sigma(K^0_S)=1.826\pm 0.006$~MeV$/c^2$ and $\sigma(K^0_SK^+K^-)=2.915\pm0.009$~MeV$/c^2$ are widths of the core Gaussian function and $\rho=0.602\pm 0.002$ is the correlation coefficient. The above uncertainties are statistical only. We define the signal box in the plane of rotated masses $m'(K^0_S)$ and $m'(K^0_SK^+K^-)$ in order to minimize correlations. Signal window in $|m'(K^0_S)|$ and $|m'(K^0_SK^+K^-)|$ is chosen to minimize the expected statistical error on $y_{CP}$, using the tuned MC: we require $|m'(K^0_S)|<3.9$ and $|m'(K^0_SK^+K^-)|<2.2$. The selection criteria on $\sigma_t$ and $K^0_S$ candidate flight distance in $r-\phi$ plane, given above, are determined in the same way. We find $139\times 10^{3}$ signal events with purity of 94\%.

The lifetime difference \zdtau\ (Eq. \ref{eq_deltatau}) is determined from \zdecay\ proper decay time distributions by measuring lifetime of signal events in ON and OFF $m(K^+K^-)$ regions. The lifetime of signal events is obtained in the following way. For each event category $i$ the proper decay time distribution $P_i(t)$  is assumed to be either exponential or a delta function, convoluted with a resolution function $R_i(t)$. The distribution for all event categories is then
\begin{equation}
 P(t)=\sum_i p_iP_i(t)\otimes R_i(t),
\end{equation}
where $p_i=N_i/\sum_jN_j$ is a fraction of the category $i$. By grouping the events into the signal and background one can also write
\begin{equation}
 P(t)=p\frac{1}{\tau_s}e^{-t/\tau_s}\otimes R_s(t)+(1-p)B(t),
 \label{eq_pdt}
\end{equation}
where the first term represents the measured distribution of a signal with lifetime $\tau_s$, $R_s(t)$ is a signal resolution function and $p=N_s/(N_s+N_b)$ is a fraction of signal events. The last term represents the distribution of background events.
The mean of the above distribution (Eq. \ref{eq_pdt}) is
\begin{equation}
 <\!t\!>=p(\tau_s+t_0)+(1-p)<\!t\!>_b,
 \label{eq_meant}
\end{equation}
where $t_0$ is the mean of the signal resolution function $R_s(t)$ and $<\!t\!>_b$ is the mean lifetime of the background. The lifetime of signal events, shifted for the resolution function offset, can be calculated from Eq. \ref{eq_meant}
\begin{equation}
 \tau_s+t_0 = \frac{<\!t\!>-(1-p)<\!t\!>_b}{p}
 \label{eq_taus}
\end{equation}
with uncertainty
\begin{equation}
 \sigma_{\tau_s}^2=\left ( \frac{1}{p}\sigma\right)^2 +\left ( \frac{1-p}{p}\sigma_b\right)^2+\left ( \frac{<\!\tau\!>-<\!\tau\!>_b}{p^2}\sigma_p\right)^2,
\end{equation}
where $\sigma$, $\sigma_b$ and $\sigma_p$ are determined from the proper decay time distributions of all events $P(t)$ and background events $B(t)$ in the following way
\[
 \sigma = \frac{rms(P)}{\sqrt{N}},~~~ \sigma_b = \frac{rms(B)}{\sqrt{N_b}}~~~\mbox{and}~~~\sigma_p = \sqrt{\frac{p(1-p)}{N}}.
\]
The $B(t)$ distribution of background events populating the signal window is approximated by the proper decay time distribution of events taken from $m'(K^0_SK^+K^-)$ sideband of equal size as signal window. No scaling factor is needed, since the background events are linearly distributed in $m'(K^0_SK^+K^-)$. The tuned MC is used to select the sideband region that best reproduces the timing distribution of background events in $m'(K^0_SK^+K^-)$ signal window, which is chosen to be $9.7 < |m'(K^0_SK^+K^-)| < 11.9$.

\begin{figure}[t!]
 \begin{center}
 \includegraphics[width=1.0\textwidth]{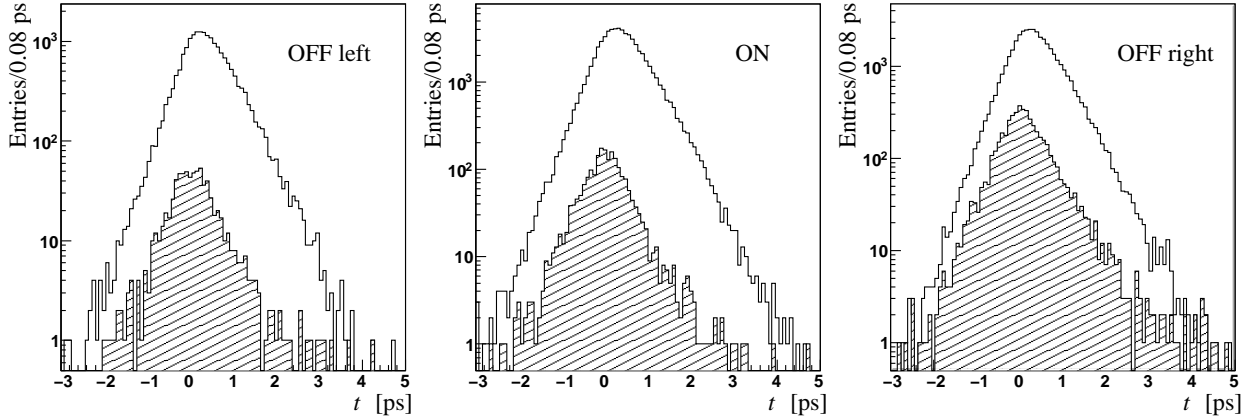}
 \end{center}
 \caption{Proper decay time distributions for events populating $m(K^+K^-)<1.010$~GeV/$c^2$ (left), $1.015<m(K^+K^-)<1.025$~GeV/$c^2$ (middle) and $1.033<m(K^+K^-)<1.100$~GeV/$c^2$ (right). The hatched area histograms show the contribution of events populating the $m'(K^0_SK^+K^-)$ sideband.}
 \label{fig_time}
\end{figure}

\begin{table}[t!]
\begin{center}
\begin{tabular}{c|ccccc|c}\hline \hline
$m(K^+K^-)$ & $N_{\rm sw}$        & $N_{\rm sb}$        & $<\!t\!>_{\rm sw}$ [fs] & $<\!t\!>_{\rm sb}$ [fs] & $p$ [\%]           & $\tau_s+t_0$~$[$fs$]$  \\ \hline
OFF left    & 19618           & 763             & $400.2\pm 4.5$      & $121.2\pm27.7$      & $96.11\pm0.14$     & $411.5\pm 4.8$          \\ 
ON          & 66112           & 2104            & $403.0\pm 2.4$      & $41.2\pm13.8$       & $96.82\pm0.07$     & $414.9\pm 2.6$         \\
OFF right   & 40634           & 4879            & $381.6\pm 3.2$      & $138.6\pm10.2$      & $87.99\pm0.16$     & $414.7\pm 3.9$         \\\hline
\end{tabular}
\end{center}
\caption{Numbers of events in the signal window $N_{\rm sw}$ and sideband $N_{\rm sb}$, mean proper decay times of events in the signal window $<\!t\!>_{\rm sw}$ and $<\!t\!>_{\rm sb}$, fraction of signal events in the signal window $p=1-N_{\rm sb}/N_{\rm sw}$ and reconstructed lifetime $\tau_s+t_0$ (Eq. \ref{eq_taus}) shifted for resolution function offset obtained on untagged real data sample.}
\label{tab_untagged_DATA}
\end{table}
In Table \ref{tab_untagged_DATA} the numbers of reconstructed events in the signal window $N_{\rm sw}$ and sideband $N_{\rm sb}$, mean proper decay times of events in the signal window $<\!t\!>_{\rm sw}$ and $<\!t\!>_{\rm sb}$, fraction of signal events in the signal window $p=1-N_{\rm sb}/N_{\rm sw}$ and reconstructed lifetime $\tau_s+t_0$ (Eq. \ref{eq_taus}) shifted for resolution function offset obtained on real data sample are given for 3 different regions: OFF left ($m(K^+K^-)<1.010$~GeV/$c^2$), ON ($1.015<m(K^+K^-)<1.025$~GeV/$c^2$) and OFF right ($1.033<m(K^+K^-)<1.100$~GeV/$c^2$). Figure \ref{fig_time} shows proper decay time distributions for events populating OFF left, ON and OFF right $m(K^+K^-)$ regions.

\begin{figure}[t!]
 \begin{center}
 \includegraphics[width=0.8\textwidth]{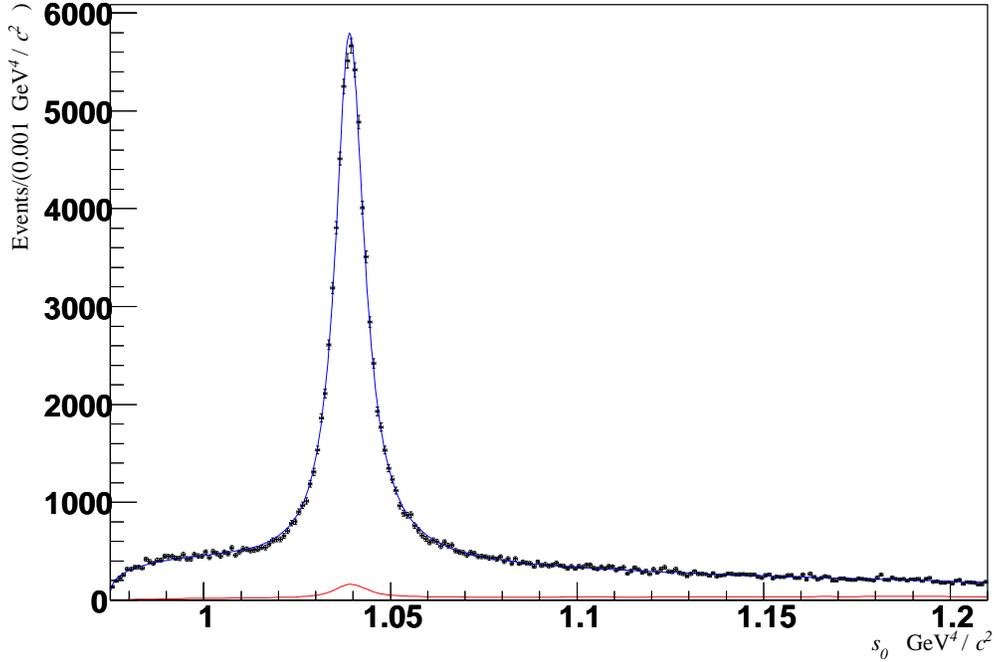}
 \end{center}
 \caption{$s_0$ distribution of \zdecay\ decays with superimposed fit results with Dalitz model given in Ref. \cite{BaBar} (right). The blue solid line is the overall fitted function and the red line is the background contribution. }
 \label{fig_s0_untaggedfits}
\end{figure}

To obtain $y_{CP}$ from measured \zdtau\ (Eq. \ref{eq_deltatau}) the fraction difference $f_1^{\rm ON} - f_1^{\rm OFF}$ is needed. Dalitz models of \zdecay\ decays given in Ref. \cite{BaBarII,BaBar} are used to fit the $s_0$ distribution. The $s_0$ distribution of signal events is parametrized as 
\begin{equation}
 {\cal P}_{\rm sig}=\varepsilon(s_0)\int \varepsilon(s_+)|{\cal M}(s_0,s_+)|^2ds_+,
\end{equation}
where $|{\cal M}(s_0,s_+)|^2$ is the time integrated decay rate (Eq. \ref{eq_mD0}), and $\varepsilon(s_0)$ ($\varepsilon(s_+)$) is the reconstruction efficiency in $s_0$ ($s_+$) determined from a sample of MC events in which the decay mode was generated according to phase space. Efficiency in $s_0$ and $s_+$ is assumed to be factorizable. No significant difference is observed between $\varepsilon(s_+)$ obtained for events populating ON and OFF $s_0$ regions. All phases, amplitudes, masses and widths of the resonances are fixed to the values determined in Ref. \cite{BaBarII,BaBar}, except for the amplitudes of $K^0_S \phi(1020)$ and $K^- a_0(980)^+$ ($K^- a_0(1450)^+$) contributions using model from Ref. \cite{BaBarII} (\cite{BaBar}). The free parameters of the fit are also the coupling constant $g_{KK}$ of a coupled channel BW \cite{BaBarII} which describes the $a_0(980)$ resonance and the mass and width of the $\phi(1020)$ resonance in order to account for mass resolution effects. To describe background events in the $s_0$ distribution, events from the  $m'(K^0_SK^+K^-)$ sideband are taken. The $\chi^2$ test of the MC $s_0$ distributions of background events taken from the signal window and sideband yields $\chi^2/ndf=136/99$. The fraction of signal events in the signal window $p=1-N_{\rm sb}/N_{\rm sw}$ is determined from the numbers of events in the signal window $N_{\rm sw}$ and sideband $N_{\rm sb}$ and it is fixed parameter of the fit. Figure \ref{fig_s0_untaggedfits} shows the fit result to the $s_0$ distribution for the Dalitz model given in Ref. \cite{BaBar}. The $\chi^2/ndf$ value of the fit is $431.8/230$ using the Dalitz model from Ref. \cite{BaBarII} and $291.7/230$ using the Dalitz model from Ref. \cite{BaBar}. In Table \ref{tab_fractionsBest} fractions $f_1^{\rm ON}$ and $f_1^{\rm OFF}$  and the fraction difference $f_1^{\rm ON}-f_1^{\rm OFF}$ are given for both Dalitz models.
Although the models are very different, with different resonant structure \cite{model_diff}, the fraction differences calculated for each model are in agreement.

\begin{table}[t!]
\begin{center}
 \begin{tabular}{c|ccc|ccc}\hline \hline
                         & \multicolumn{3}{c|}{Nominal}                        & \multicolumn{3}{c}{Fitted}\\ \hline
 Model                   & $f_1^{\rm ON}$  & $f_1^{\rm OFF}$ & $f_1^{\rm ON} - f_1^{\rm OFF}$  & $f_1^{\rm ON}$  & $f_1^{\rm OFF}$ & $f_1^{\rm ON} - f_1^{\rm OFF}$\\
 \hline
 4 res. \cite{BaBarII}   & 0.117       & 0.847       & $-0.730\pm0.031$       & 0.113       & 0.844       & $-0.732\pm0.003$\\ 
 8 res. \cite{BaBar}     & 0.124       & 0.877       & $-0.753\pm0.004$       & 0.111       & 0.880       & $-0.769\pm0.005$\\
 \hline
\end{tabular}
\end{center}
\caption{Fractions $f_1^{\rm ON}$ and $f_1^{\rm OFF}$ ($f^{ON/OFF}_1 = \oint_{\rm ON/OFF} |{\cal A}_1|^2/\oint_{\rm ON/OFF} (|{\cal A}_1|^2+|{\cal A}_2|^2$) and the fraction difference $f_1^{\rm ON}-f_1^{\rm OFF}$ for the two Dalitz models Ref. \cite{BaBarII,BaBar}. The nominal values are calculated using the given Dalitz models in Ref. \cite{BaBarII,BaBar} and fitted values using the obtained values of free parameters of the fit to the $s_0$ distribution. Uncertainties on $f_1^{\rm ON}-f_1^{\rm OFF}$ were calculated using the statistical errors of amplitudes and phases given for each model, without taking into account any correlation between the amplitudes and phases. }
\label{tab_fractionsBest}
\end{table}

The reconstructed lifetimes shifted for the resolution function offset, $\tau_s + t_0$, of $D^0$ candidates in ON and OFF regions are $414.9\pm2.6$ fs and $413.6\pm3.1$ fs, respectively, from which $\Delta_{\tau}=(-0.16\pm0.48)\%$ is obtained. We assumed that the resolution function offset, $t_0$, is equal for the events populating the ON and OFF regions and much smaller than $D^0$ lifetime. Using the Eq. \ref{eq_deltatau} and the fraction difference $f_1^{\rm ON} - f_1^{\rm OFF}=-0.769$, obtained by fitting $s_0$ distribution with Dalitz model given in Ref. \cite{BaBar}, yields $y_{CP}=(0.21\pm0.63(\rm stat.))\%$. 

We consider systematic uncertainties arising from both experimental sources and from the \zdecay\ model. First, we check on the MC sample if the resolution function offsets, $t_0^{\rm ON}$ and $t_0^{\rm OFF}$ are equal. They are in agreement within the statistical uncertainty and small $(t_0=0.7\%\cdot\tau_{D^0})$. Next, we vary the sideband in $m'(K^0_SK^+K^-)$ used to describe the background populating the signal window and measure for each sideband the \zdtau. For different sidebands used the obtained \zdtau\ values are in agreement. The maximal difference in \zdtau\ was taken to estimate the systematic uncertainty. Finally, possible systematic effects of selection criteria were studied by varying the signal box sizes, and cut values on $\sigma_t$ and $K^0_S$ flight distance in $r-\phi$ plane. Again no statistical significant deviation was observed and the maximal difference in \zdtau\ was taken to estimate the systematic uncertainty. We add all different sources in quadrature to obtain the overall experimental systematic uncertainty summarized in Table \ref{tab_syst}.
\begin{table}[t!]
\begin{tabular}
{@{\hspace{0.5cm}}l@{\hspace{0.5cm}}|@{\hspace{0.5cm}}c@{\hspace{0.5cm}}}
\hline \hline
Source & Systematic error (\%) \\
\hline
Resolution function offset difference $t^{\rm OFF}_0-t^{\rm ON}_0$ & $\pm 0.21$ \\
Selection of $m'(K^0_SK^+K^-)$ sideband                    & $\pm 0.35$ \\
Variation of selection criteria                            & $\pm 0.44$ \\
\hline
Total & $\pm 0.60$ \\
\hline 
\end{tabular}
\caption{Sources of the systematic uncertainty for \zdtau.}
\label{tab_syst}
\end{table}

The systematic uncertainty due to our choice of \zdecay\ decay model is evaluated as follows. First, we compare the fraction difference $f_1^{\rm ON} - f_1^{\rm OFF}$ obtained using the Dalitz Models in Ref. \cite{BaBarII,BaBar}. Despite the differences between the two models in terms of the resonant structure \cite{model_diff}, the fraction differences  $f_1^{\rm ON} - f_1^{\rm OFF}$ (given in Tab. \ref{tab_fractionsBest}) are in agreement. We assign 3\% relative error for measured $y_{CP}$ due to small difference in the above fractions. An additional 2\% relative error for measured $y_{CP}$ is assigned due to the small difference between fitted and nominal values of fraction difference $f_1^{\rm ON} - f_1^{\rm OFF}$ (given in Tab. \ref{tab_fractionsBest}). The real and imaginary part of the interference term ${\cal A}_1{\cal A}_2^{\ast}$ in the decay rate (Eq. \ref{eq_mD0}) are zero after integrating over the $s_+$. Since the reconstruction efficiency is not constant in $s_+$, this is not entirely true. However, even if the observed $s_+$ reconstruction efficiency is taken into account this has negligible effect and Eq. \ref{eq_deltatau} still holds. This was also verified by MC with non-zero $x$ and $y$ values of mixing parameters, where the detector response was simply simulated by randomly rejecting events according to the observed dependence of efficiency in $s_+$. The difference between the obtained $\Delta_{\tau}$ values (with and without taking into account the efficiency in $s_+$) are in agreement within statistical uncertainty, so no additional systematical uncertainty is assigned. Adding all variations in quadrature, the obtained relative model systematic uncertainty is 4\%.

In summary, we determine \ycp\ by measuring the difference in lifetimes between $D^0$ mesons decaying to $K^0_SK^+K^-$ in two different $m(K^+K^-)$ regions with different contributions of $CP$ even and odd eigenstates to be
\[
 y_{CP} = (0.21\pm 0.63 ({\rm stat.})\pm 0.78 (\rm syst.) \pm 0.01(\rm model))\%.
\]
The result is in agreement with world average of $y_{CP}$ of previous measurements \cite{hfag,Schwartz}.

We thank the KEKB group for the excellent operation of the
accelerator, the KEK cryogenics group for the efficient
operation of the solenoid, and the KEK computer group and
the National Institute of Informatics for valuable computing
and SINET3 network support. We acknowledge support from
the Ministry of Education, Culture, Sports, Science, and
Technology of Japan and the Japan Society for the Promotion
of Science; the Australian Research Council and the
Australian Department of Education, Science and Training;
the National Natural Science Foundation of China under
contract No.~10575109 and 10775142; the Department of
Science and Technology of India; 
the BK21 program of the Ministry of Education of Korea, 
the CHEP SRC program and Basic Research program 
(grant No.~R01-2005-000-10089-0) of the Korea Science and
Engineering Foundation, and the Pure Basic Research Group 
program of the Korea Research Foundation; 
the Polish State Committee for Scientific Research; 
the Ministry of Education and Science of the Russian
Federation and the Russian Federal Agency for Atomic Energy;
the Slovenian Research Agency;  the Swiss
National Science Foundation; the National Science Council
and the Ministry of Education of Taiwan; and the U.S.\
Department of Energy.


\begin{thebibliography}{99}
\bibitem{CC}
Throughout this paper, the inclusion of the charge conjugate mode decay is implied unless otherwise stated.
\bibitem{vertex}
If both charged kaons are used to determine the $D^0$ candidate's decay point a reconstructed proper time and $m(K^+K^-)$ correlation is observed around nominal mass of the $\phi(1020)$.
\bibitem{model_diff}
In the Dalitz analysis of \zdecay\ decays \cite{BaBarII} the Dalitz model consists of $K^0_S a_0(980)^0$, $K^0_S \phi(1020)$, $K^0_S f_0(1370)$, $K^0_S f_0(980)$, and $K^- a_0(980)^+$ contributions. The fitted fractions of the latter two is consistent with 0 and authors do not quote their amplitudes and phases, so these to contributions are not used in this paper. In the Dalitz analysis \cite{BaBar} the Dalitz model consists of  $K^0_S a_0(980)^0$, $K^0_S \phi(1020)$, $K^0_S f_0(1370)$, $K^0_S f_2(1270)$, $K^0_S a_0(1450)^0$, $K^- a_0(980)^+$, $K^- a_0(1450)^+$ and $K^+ a_0(980)^-$.
\end{thebibliography}
\end{document}